\documentclass[twocolumn,prl,showpacs]{revtex4} 
\usepackage{graphicx,amssymb,amsmath,bm}

\begin{document}
 
\title {Magnetic couplings, optical spectra, and spin-orbit exciton in 
$5d$ electron Mott insulator Sr$_2$IrO$_4$}

\author {Beom Hyun Kim$^1$} 
\author {G. Khaliullin$^2$}
\author {B. I. Min$^1$}
\email[]{bimin@postech.ac.kr}
\affiliation{$^1$Department of Physics, PCTP, 
Pohang University of Science and Technology, Pohang 790-784, Korea}
\affiliation{$^2$Max Planck Institute for Solid State Research, 
Heisenbergstrasse 1, D-70569 Stuttgart, Germany}
\date{\today}

\begin{abstract}
Based on the microscopic model including spin-orbit coupling, on-site Coulomb 
and Hund's interactions, as well as crystal field effects, we have 
investigated magnetic and optical properties of Sr$_2$IrO$_4$. Taking into 
account all intermediate state multiplets generated by virtual hoppings of 
electrons, we calculated the isotropic, pseudodipolar, and 
Dzyaloshinsky-Moriya coupling constants, which describe the experiment quite 
well. The optical conductivity $\sigma(\omega)$ evaluated by the exact 
diagonalization method shows two peaks at $\sim 0.5$ and $\sim 1.0$ eV 
in agreement with experiment. The two peak structure of $\sigma(\omega)$ 
arises from the unusual Fano-type overlap between electron-hole continuum 
of the $J_{eff}=1/2$ band and the intrasite spin-orbit exciton observed 
recently in Sr$_2$IrO$_4$.   
\end{abstract}


\maketitle


Mott physics is one of the most fundamental phenomena in condensed matter 
physics giving rise to diverse and fascinating collective behavior of
correlated electrons~\cite{Ima98}. In Mott insulators, strong on-site Coulomb 
repulsion ($U$) splits the half-filled band into the lower Hubbard band (LHB) 
accommodating the spin and orbital degrees of freedom of electrons, and the 
empty upper Hubbard bands (UHB). The electrons hop into adjacent sites only 
virtually, overcoming thereby the Mott-Hubbard gaps. Typically, the 
antiferromagnetic (AFM) ground state is realized due to the kinetic energy 
gain of the virtual exchange processes. The optical conductivity shows peak 
structures at the absorption edges corresponding to the transitions between 
LHB and UHBs.
 
Recently, a new class of $5d$ Mott insulators such as 
Sr$_2$IrO$_4$~\cite{BJKim1,SJMoon,BJKim2} and Na$_2$IrO$_3$~\cite{Singh,Tak00} 
has been discovered, where the strong spin-orbit (SO) coupling is crucial 
for stabilizing the insulating state. The Coulomb interaction between 
$5d$-electrons $U \sim 2$ eV is much smaller than that in conventional $3d$ 
electron Mott insulators. On the other hand, the SO coupling 
$\lambda \simeq 0.4$ eV in iridates is much larger and splits $^{2}T_{1g}$ 
states of $5d^5$-shell into the half-filled $J_{eff}=1/2$ and fully-occupied 
$J_{eff}=3/2$ states. Then the narrow $J_{eff}=1/2$ band tends to undergo Mott 
transition even at relatively small 
$U$~\cite{BJKim1,SJMoon,Watanabe,Martins,Arita}.

The issue in iridates is to which extent the physical properties and model 
descriptions of $J_{eff}=1/2$ systems are (dis)similar to those of
conventional $3d$ Mott insulators. The best studied member of $5d$ Mott systems 
is Sr$_2$IrO$_4$ perovskite. Its in-plane canted AFM ground 
state~\cite{BJKim2}, magnon spectra~\cite{JKim}, and finite temperature spin 
dynamics~\cite{Fuj12} closely resemble those of parent high-T$_c$ cuprates, 
in accord with the theoretical predictions~\cite{Jackeli} based on 
Mott-Hubbard picture. In addition to magnons, resonant inelastic x-ray 
scattering  (RIXS) experiments~\cite{JKim,Ishii} have observed also the higher 
energy broad peak at $\sim 0.5-0.8$~eV. Based on the theoretical 
expectations~\cite{Jackeli,Kha04,Ament} that the SO split $t_{2g}$ 
manifold should have a magnetically active mode at $\sim \frac{3}{2}\lambda$, 
this peak has been attributed to the transition between $J_{eff}=1/2$ and   
$3/2$ states and termed ``SO exciton''~\cite{JKim}.  

Concerning the charge excitation spectra, the optical conductivity 
$\sigma(\omega)$ of Sr$_2$IrO$_4$ shows two peaks at $\sim 0.5$ and 
$\sim 1.0$ eV~\cite{BJKim1} that are preserved up to high 
temperatures~\cite{SJMoon2}. The first peak was assigned to the transition 
from occupied $J_{eff}=1/2$ LHB to $J_{eff}=1/2$ UHB, while the peak at 
$\sim 1.0$~eV to that from $J_{eff}=3/2$ to unoccupied 
$J_{eff}=1/2$~\cite{BJKim1}. This interpretation, however, is based on a
picture of single-electron density of states. In fact, there has been no 
theoretical calculations of $\sigma(\omega)$ in iridates taking into account 
the many-electron multiplet structure of excited states, which is known to be
essential for the interpretation of optical data. Moreover, there are inherent 
relations between the optical absorption peaks/intensities and the strength 
of magnetic couplings (both are determined by the same virtual hoppings and 
excited multiplets), which enable one to extract the physical parameters, 
such as $U$ and Hund's coupling $J_H$, from a combined analysis of the 
magnetic and optical data~\cite{Kov04,Khaliullin,Kha05}. The aim of the 
present Letter is to extend this fruitful approach to the $5d$ electron 
Mott insulators. 

We have calculated the magnetic couplings, optical conductivity, and RIXS 
spectra in Sr$_2$IrO$_4$ by exact diagonalization (ED) of a microscopic model 
on small clusters, fully incorporating the multiplet structure of Ir ions, SO 
coupling, tetragonal distortion, and octahedral rotations. The magnetic 
couplings obtained are consistent with the available 
data~\cite{BJKim2,JKim,Fuj12}. Calculated RIXS spectra reproduce the SO 
exciton mode. More interestingly, we found that, unlike the case of $3d$ 
oxides~\cite{Kov04,Khaliullin,Kha05}, the observed peaks in $\sigma(\omega)$ 
of Sr$_2$IrO$_4$ cannot be directly determined from the multiplet energies 
and intensities. Instead, the strong mixing between intersite optical 
excitations (electron-hole continuum) and intraionic transition between 
$J_{eff}=1/2$ and $3/2$ states (SO exciton) is found to be essential feature 
of Sr$_2$IrO$_4$. This implies that neither purely atomic nor simple band 
picture is sufficient to describe the small charge-gap iridium oxides. 

\begin{figure}[t]
\centering
\includegraphics[width=7.6 cm]{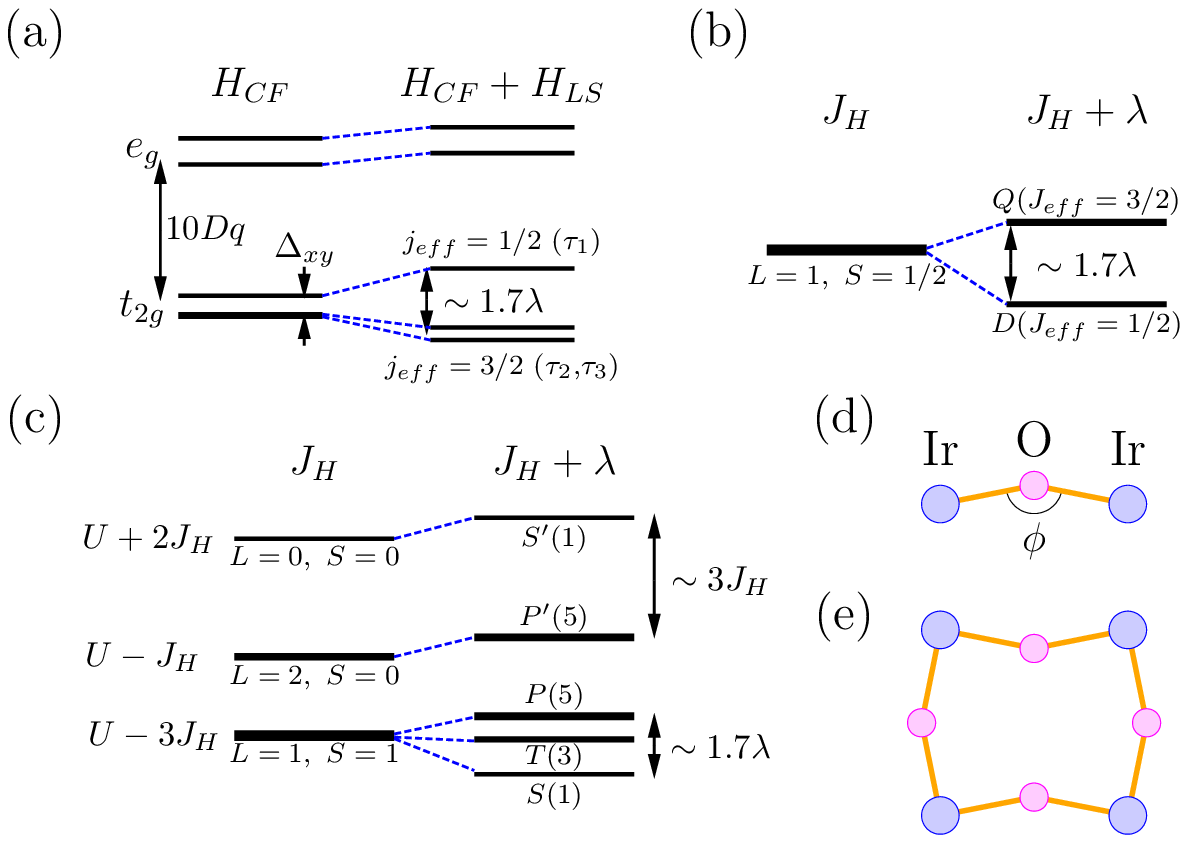}
\caption
{ (color online)
(a) Energy levels of $d$ electron in the presence of the cubic crystal 
field $10Dq$, the tetragonal crystal field $\Delta_{xy}$, and the SO 
interaction $\lambda$~\cite{para1}. The lowest multiplet levels of 
(b) $d^5$ and (c) $d^4$ configurations including SO $\lambda$ and the Hund's 
coupling $J_H$. 
(d) Two-site and (e) four-site clusters in the $xy$-plane, which were employed 
in the evaluation of the magnetic couplings and conductivity $\sigma(\omega)$.
}
\label{fig1}
\end{figure}

{\it Model.}-- To describe the electronic structure of Ir ions, we adopted 
the following Hamiltonian:
\begin{align}
\label{eq1}
&H_{ion} =\sum_{\tau\tilde{\sigma}} \epsilon_{\tau} n_{\tau\tilde{\sigma}} 
   + \frac{1}{2}\sum_{\sigma\sigma^{\prime}\mu\nu} 
  U_{\mu\nu} c_{\mu\sigma}^{\dagger}c_{\nu\sigma^{\prime}}^{\dagger}
  c_{\nu\sigma^{\prime}}c_{\mu\sigma}  \\ 
  &+ \frac{1}{2}\sum_{\substack{\sigma \sigma^{\prime} \\ \mu\ne\nu}}
  J_{\mu\nu}c_{\mu\sigma}^{\dagger}c_{\nu\sigma^{\prime}}^{\dagger}
  c_{\mu\sigma^{\prime}}c_{\nu\sigma} 
   + \frac{1}{2}\sum_{\substack{\sigma \\ \mu\ne\nu}} 
  J_{\mu\nu}^{\prime} c_{\mu\sigma}^{\dagger}c_{\mu-\sigma}^{\dagger}
  c_{\nu-\sigma}c_{\nu\sigma},  \nonumber 
\end{align}
where $\tau$ and isospin $\tilde{\sigma}$~\cite{Jackeli} refer to the lowest 
three Kramers doublets [the eigenstates in the presence of the tetragonal 
crystal field splitting $\Delta_{xy}$ and SO coupling, see Fig.~\ref{fig1}(a)]. 
The other terms describe the on-site Coulomb and Hund's interactions, where 
$\mu$ and $\sigma$ are the orbital and spin indices, respectively. 
As usual, we parametrize the interaction matrix as $U_{\mu\mu}=U$, 
$U_{\mu\ne\nu}=U-2J_H$, and $J_{\mu\nu}=J_{\mu\nu}^{\prime}=J_H$~\cite{Kan63}.

Fig.~\ref{fig1}(b,c) shows all possible multiplets of $d^5$ and $d^4$ 
electronic configurations. For $d^6$, all three lowest Kramers doublets 
$\tau_{1,2,3}$ are fully occupied. We note that the wave-functions of these 
states are often taken in the limit of $10Dq\rightarrow\infty$, i.e., 
admixture of $e_g$ orbitals ($\propto\lambda/10Dq$) in the ground state via 
SO coupling is ignored. Since the SO coupling in iridates is strong, we fully 
include the $e_g$ orbital admixture in the lowest Kramers doublets. 

Because an electron hops between nearest-neighbor (NN) Ir sites via an oxygen 
[see Fig.~\ref{fig1}(d)], and the charge transfer energy $\Delta$ between
Ir-5d and O-2p is large ($\Delta \sim 3.3$ eV~\cite{SJMoon3}), we assumed 
the following effective NN hopping Hamiltonian:
\begin{equation}
 H_{ij}=\sum_{\tau\tau^{\prime}\tilde{\sigma}\tilde{\sigma}^{\prime}} 
 \left(t_{i\tau\tilde{\sigma};j\tau^{\prime}\tilde{\sigma}^{\prime}}
 c_{j\tau^{\prime}\tilde{\sigma}^{\prime}}^{\dagger}
 c_{i\tau\tilde{\sigma}} + h.c.\right),
\label{eq2}
\end{equation}
where the effective hopping integral between $\tau\tilde{\sigma}$ 
and $\tau^{\prime}\tilde{\sigma}^{\prime}$ states is
$t_{ i\tau_i\tilde{\sigma};j\tau^{\prime}\tilde{\sigma}^{\prime}}
=\sum_{p\sigma}\frac{t_{i\tau\tilde{\sigma};p\sigma}
 t^*_{j\tau^{\prime}\tilde{\sigma}^{\prime};p\sigma}} 
{\sqrt{(\Delta+\epsilon_{\tau})(\Delta+\epsilon_{\tau^{\prime}})}}$.
We calculated the $pd$-hopping matrix $t_{i\tau\tilde{\sigma};p\sigma}$ between 
5$d$ Ir and 2$p$ O orbitals in terms of two parameters $t_{pd\sigma}$ and 
$t_{pd\pi}$~\cite{Slater}. 

\begin{table}[b]
\caption
{Physical parameters in units of eV.} \label{Para}
\begin{ruledtabular}
\begin{tabular}{c c c c c c c c }
 $10Dq$ & $\Delta$ & $\Delta_{xy}$ & $U$ & $J_H$ & $\lambda$ & 
 $t_{pd\sigma}$ & $t_{pd\pi}$ \\
\hline
3.0 & 3.3 & 0.15 & 1.86 & 0.5 & 0.4 & $-$1.8& 0.83 \\
\end{tabular}
\end{ruledtabular}
\end{table}

{\it Magnetic interactions.}-- 
We considered the Ir$_1$-Ir$_2$ pair described by the following Hamiltonian:
\begin{equation}
H = H_1 + H_2 + H_{12}.
\label{eq3}
\end{equation}
Here $H_1$ and $H_2$ are the ionic Hamiltonians given by Eq.~(\ref{eq1}), 
and $H_{12}$ is the hopping term of Eq.~(\ref{eq2}). Using the ED method, we 
solved this Hamiltonian numerically, taking into account all possible 
multiplets allowed for the $d^5-d^5$ and $d^4-d^6$ configurations, and 
obtained eigenvalues $E_n$ and eigenstates $|\psi_n\rangle$ of the Ir-Ir 
cluster. The Hamiltonian can be then expressed as 
$H\!=\!\sum_{n}E_n|\psi_n\rangle\langle \psi_n|$.
Without the hopping term $H_{12}$, the Ir-Ir pair would have four degenerate 
states denoted by $|\psi_i^0 \rangle$ ($i=1$-$4$). The magnetic interactions 
are generated by virtual hoppings among them. The resulting effective 
magnetic Hamiltonian can be obtained by applying the projection operator
$\mathcal{P}_{1/2}\!=\!\sum_{i=1}^{4}| \psi_{i}^{0}\rangle\langle \psi_{i}^{0}|$
onto $H$~\cite{BHKim}:
\begin{equation}
\mathcal{P}_{1/2}H\mathcal{P}_{1/2}-
 \frac{1}{4} \textrm{Tr}\left(\mathcal{P}_{1/2}H\mathcal{P}_{1/2}\right)
 = \tilde{\mathbf{S}}_1\cdot\mathcal{J}\cdot\tilde{\mathbf{S}}_2,
\label{eqj}
\end{equation}
where $\tilde{\mathbf{S}}$ is isospin one-half, and $\mathcal{J}$ 
is a $3\times3$ tensor. 

Consistent with symmetry considerations as well as with Ref.~\cite{Jackeli}, 
we found that Eq.~(\ref{eqj}) comprises four distinct nonvanishing terms,  
\begin{equation}
\label{Jform}
J\tilde{\mathbf{S}}_{1}\!\cdot\tilde{\mathbf{S}}_{2}
+\delta\!J_{z} \tilde{S}_{1,z} \tilde{S}_{2,z}
+\delta\!J_{xy} (\tilde{\mathbf{S}}_{1}\!\cdot \vec{r}_{12}) 
          (\tilde{\mathbf{S}}_{2}\!\cdot \vec{r}_{12})
 +\mathbf{D}\cdot \tilde{\mathbf{S}}_{1}\!\times \tilde{\mathbf{S}}_{2},
\end{equation}
corresponding to isotropic Heisenberg ($J$), symmetric ($\delta J_{z}$, 
$\delta J_{xy}$) and antisymmetric ($\mathbf{D}$) anisotropic couplings 
between NN Ir-moments. We note that only the $z$-component 
Dzyaloshinsky-Moriya (DM) vector $\mathbf{D}$ survives because of the 
mirror symmetry with respect to the $xy$-plane, i.e., $\mathbf{D}=(0,0,D)$. 

\begin{figure}[t]
\centering
\includegraphics[width=7.6 cm]{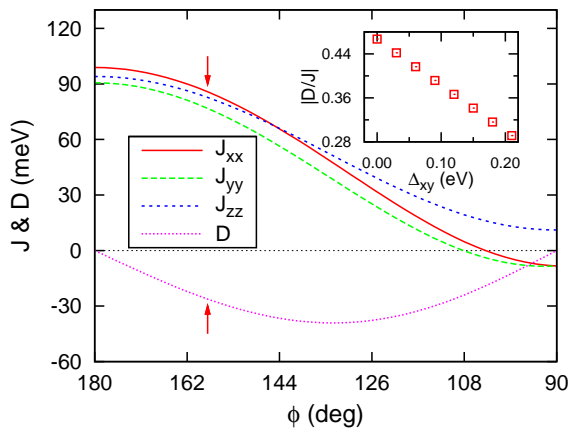}
\caption
{(color online)
Magnetic coupling constants between NN Ir ions ($\vec{r}_{12}||x$) with 
respect to the bonding angle ($\phi$). $J_{xx}$, $J_{yy}$, and $J_{zz}$ 
represent diagonal parts of the superexchange tensor 
$\mathcal{J}$. $J_{xx}=J+\delta J_{xy}$, $J_{yy}=J$, and $J_{zz}=J+\delta J_z$ 
are deduced from Eq.~(\ref{Jform}). $D$ is a $z$-component of the DM 
vector $\mathbf{D}$. Arrows denote the bonding angle in Sr$_2$IrO$_4$. 
Inset shows $|D/J|$ ratio as a function of the tetragonal splitting 
$\Delta_{xy}$.
}
\label{fig2}
\end{figure}

Figure~\ref{fig2} presents the magnetic coupling constants, calculated by using 
the parameters provided in Table~\ref{Para}, as a function of 
Ir-O-Ir bonding angle $\phi$. $10Dq$ and $\Delta$ values are adopted from 
Refs.~\cite{SJMoon3,Ishii,Kat12} and $\lambda$ from Refs.~\cite{Sch84,Kat12}.
We have determined the optimal values of $U$, $J_H$, $t_{pd}$ such that they 
describe both the magnetic and optical data properly~\cite{SM}. We set tetragonal 
splitting $\Delta_{xy}=0.15$ eV; at this value, $|D/J|$ ratio becomes 
$\simeq 0.34$ and yields the spin canting angle of $\simeq 9.3^{\circ}$ 
at $\phi=158^{\circ}$ (as in Sr$_2$IrO$_4$). As shown in the inset 
of Fig.~\ref{fig2}, when $\Delta_{xy}$ is too small, $|D/J|$ ratio becomes 
too large to describe the observed spin canting. Approximately, we find 
that $D \propto \sin\left(2\phi\right)$. The spin canting angle of 
$\sim 9.3^{\circ}$ is close to those found in previous 
studies~\cite{Jackeli,HJin}. Noteworthy is that the Ising coupling
$\delta J_{z}$ ($=J_{zz}-J_{yy}$) in Eq.~(\ref{Jform}) is enhanced when decreasing
bonding angle $\phi$. We also calculated a pseudodipolar coupling  
$\delta J_{xy}$ for different $J_H$ values and found that $\delta J_{xy}$ 
becomes zero at $J_H=0$. Both of these observations agree with the
analytical results~\cite{Jackeli}.

Summarizing our results for magnetic interactions in Sr$_2$IrO$_4$, we 
obtained the following values at $\phi=158^{\circ}$: $J\simeq 76.8$, 
$D\simeq -26.2$, $\delta J_{z}\simeq 5.9$, and $\delta J_{xy}\simeq 8.6$ meV.
With these coupling constants, the interactions of Eq.~(\ref{Jform}) lead to 
the canted AFM state with Ir-moments lying in the $xy$-plane, as observed in 
Sr$_2$IrO$_4$~\cite{BJKim2}. The calculated isotropic coupling $J$ is in 
close agreement with the experimental value of $J\simeq 60$ meV~\cite{JKim}. 
Unusually large anisotropic couplings $D,\delta J_{z},\delta J_{xy}$ found 
here are the direct fingerprints of strong SO interaction.

{\it Optical conductivity and RIXS spectra.--} In $3d$ Mott insulators, the 
hopping between NN sites plays a dominant role in $\sigma(\omega)$ that 
shows peak structures near the ionic multiplet states 
$d_i^{n-1}-d_j^{n+1}$~\cite{Kov04,Khaliullin,Kha05}, and 2-site cluster may
capture essential features of $\sigma(\omega)$. In contrast, $5d$ Mott 
insulators have weaker Coulomb repulsion and thus a duality of atomic and band 
nature of correlated electrons is more pronounced~\cite{JKim}. In order to 
capture the delocalization of optically excited electron-hole (e-h) pairs, we 
consider here a $2\times2$ cluster shown in Fig.~\ref{fig1}(e).
\begin{figure}[t]
\centering
\includegraphics[width=7.0 cm]{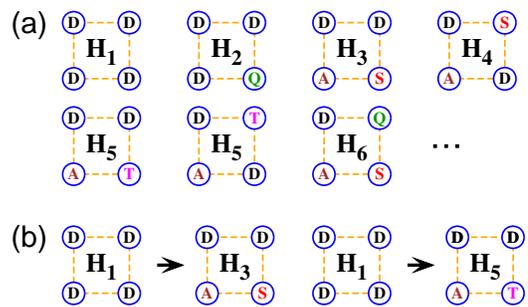}
\caption
{ (color online)
(a) Schematic diagrams of possible multiplets included in subsets 
($\mathbf{H}_i$) of the Hilbert space. $D$, $Q$, $S$, and $T$ refer to 
multiplets of $d^5$ and $d^4$, as labeled in Fig.~\ref{fig1}(b) and (c),
and $A$ denotes a nondegenerate $d^6$ state. $\mathbf{H}_5$ includes also 
diagrams (not shown) with $P, P^{\prime}, S^{\prime}$ multiplets of the $d^4$ 
state [Fig.~\ref{fig1}(c)] instead of $T$.
(b) Two examples of optically active transitions that contribute to 
$\sigma(\omega)$ at low energies. The final states $\mathbf{H}_3$ and 
$\mathbf{H}_5$ have an overlap with on-site local SO exciton $Q$ 
(i.e., with $\mathbf{H}_2$ sector), due to intersite hoppings between 
$J_{eff}=1/2$ and $J_{eff}=3/2$ states, resulting in two peak structure 
of $\sigma(\omega)$. 
}
\label{fig3}
\end{figure}

We considered all possible multiplets within the $d^5$-$d^5$-$d^5$-$d^5$ and 
$d^4$-$d^6$-$d^5$-$d^5$ charge configurations. In order to clarify the origin 
of optical peaks, it is useful to classify the Hilbert space into 
6 subspaces [see Fig.~\ref{fig3}(a)]: 
$\mathbf{H}_1$ of four $d^5$ doublets ($D$),
$\mathbf{H}_2$ of one or more quartets ($Q$) among four $d^5$ configurations,
$\mathbf{H}_3$ of two $D$'s of $d^5$ and NN e-h pairs $d^4(S)$-$d^6(A)$,
$\mathbf{H}_4$ of two $D$'s of $d^5$ and next nearest-neighbor (NNN) e-h pairs 
$d^4(S)$-$d^6(A)$, $\mathbf{H}_5$ of two $D$'s of $d^5$ and (NN \& NNN) e-h 
pairs $d^4(T,P,P^{\prime},S^{\prime})$-$d^6(A)$, and the remaining states 
($\mathbf{H}_6,...$) that involve, \textit{e.g.}, simultaneous intersite 
e-h transitions and local SO exciton $Q$. All the above configurations 
couple to each other via the hopping Hamiltonian of Eq.~(\ref{eq2}). 

We have solved the Hamiltonian matrix with the ED method, and obtained
$\sigma(\omega)$ from the following relation:
\begin{equation}
\sigma(\omega) = \pi v \frac{1\!-\!e^{-\beta \omega}}{\omega}
  \!\sum_{n<m}p_n |\langle \psi_m|\hat J_c|\psi_n\rangle|^2 
  \delta (\omega+E_n\!-\!E_m),
\label{eq5}
\end{equation}
where $v$ is volume per Ir-site, $p_n$ is the probability density of 
eigenstate $|\psi_n\rangle$, and $\hat J_c$ is the current operator. We set 
$\beta^{-1}=k_BT=30~\textrm{meV}$ to avoid the finite size effect. A function 
$\delta(\omega)$ is treated with broadening of $0.05$ eV.

\begin{figure}[t]
\centering
\includegraphics[width=7.3 cm]{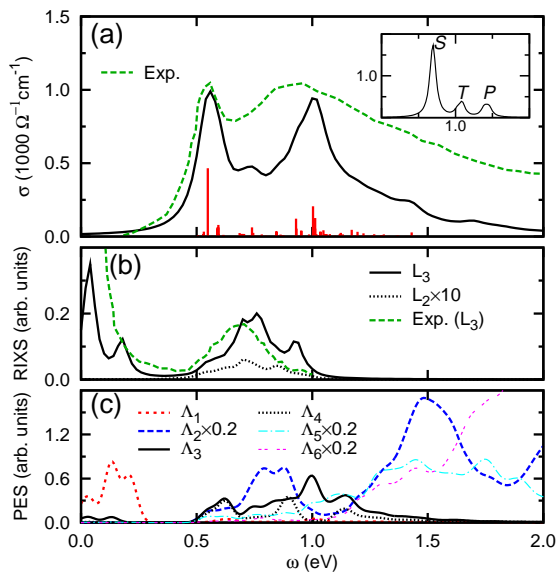}
\caption
{ (color online)
(a) Optical conductivity calculated (solid line) and experimental 
data~\cite{BJKim1} (dashed line). Red vertical sticks show relative strengths 
and positions of optical transitions without broadening. Inset: the result 
for $\sigma(\omega)$ when the engenstates $|\psi_n\rangle$ and 
energies $E_n$ in Eq.~(\ref{eq5}) are approximated by purely local ionic 
multiplets. 
(b) Calculated $L_3$-edge (solid line) and $10\times L_2$-edge (dotted line)
RIXS spectra at $\mathbf{q}=(\pi,\pi)$. Dashed line: experimental 
data~\cite{JKim}.  
(c) Projected excitation spectra: $\Lambda_1$ represents magnon band, 
$\Lambda_2$ contains one or more SO excitons, $\Lambda_3$ and $\Lambda_4$ 
represent NN and more distantly separated e-h excitations derived from the 
$J_{eff}=1/2$ states, respectively, and $\Lambda_5$ shows the e-h continuum 
of $J_{eff}=3/2$ states. $\Lambda_6$ refers to other excitations, 
e.g., simultaneous transitions in e-h and SO exciton channels. 
}
\label{fig4}
\end{figure}

Figure~\ref{fig4}(a) shows the result for $\sigma(\omega)$ calculated  
by using the parameters from Table~\ref{Para}. Two peaks are revealed 
clearly at $\sim 0.5$ and $\sim 1.0$ eV, in good agreement with experiment. 
Estimated value of $\sigma(\omega)$ at $\sim 0.5$ eV is also consistent with 
the experimental data. For $\omega>0.6$ eV, calculated $\sigma(\omega)$ is 
smaller than that observed. The possible reasons for this discrepancy are 
the contributions from the $pd$-charge transfer peak at $\sim 3$ eV and from 
two or more e-h pair excitations not included here.

Shown in Figure~\ref{fig4}(b) is the RIXS spectra, calculated by employing   
the inelastic x-ray scattering operator of Ref.~\cite{Ament} instead of 
$\hat J_c$ in Eq.~(\ref{eq5}). At the $L_3$-edge, intense magnon 
(below 0.25 eV) and SO exciton ($0.5-1.0$ eV) bands are obtained, while the 
intensity at the $L_2$-edge is nearly vanishing, in agreement with 
experiments~\cite{JKim,Ishii}. 

Of particular interest is the origin of two peaks in $\sigma(\omega)$ and 
their relation to SO exciton. In $3d$ Mott insulators, each peak would 
correspond to the specific multiplet of $d^{n-1}-d^{n+1}$ with the spectral 
weight proportional to 
$|\langle d^{n-1}d^{n+1}|J_c|d^nd^n\rangle|^2$\cite{Kov04,Khaliullin,Kha05}. 
Such a simple ionic multiplet picture, however, is incomplete in iridates. 
Indeed, as shown in inset of Fig.~\ref{fig4}(a), $\sigma(\omega)$ based on 
this picture shows that the first peak [$|DDDD\rangle \to |ASDD\rangle$ 
transition in Fig.~\ref{fig3}(b)] is much stronger than those corresponding 
to higher energy transitions involving $T,P$-multiplets~\cite{footnote}, 
in contrast to the observation that the spectral weight of $\sim 1.0$ eV peak 
is even larger than that of $\sim 0.5$ eV peak [Fig.~\ref{fig4}(a)]. 

To understand how the e-h delocalization effects lead to strong deviations
from a simple ionic picture, it is instructive to analyze the underlying 
excitation spectra in more detail. To this end, we evaluated the projected 
excitation spectrum (PES) of the $2\times2$ cluster into the subspaces  
$\mathbf{H}_{i}$ introduced above: 
\begin{equation}
\Lambda_i(\omega) = \sum_n \sum_{m\in \mathbf{H}_i} 
 |\langle \psi_n | m\rangle|^2 \delta(\omega-E_n),
\end{equation}
where $|m\rangle$ represents the orthonormal basis of the subspace
$\mathbf{H}_i$. We can identify the PES in Fig.~\ref{fig4}(c) as follows:  
$\Lambda_1$ represents magnon sector, $\Lambda_2$ is related to one or more 
intra-site $D-Q$ transitions (SO excitons), $\Lambda_3$ and $\Lambda_4$ 
are the e-h continuum of $J_{eff}=1/2$ states, and $\Lambda_5$ describes 
the e-h continuum of $J_{eff}=3/2$ bands.

We notice intriguing features in Fig.~\ref{fig4}(c):
(i) $\Lambda_1$ and $\Lambda_2$ show the peaks in the range of $0-0.25$ and 
$0.5-1.0$ eV, respectively. These peaks are manifested in the RIXS spectra as
the magnon and SO exciton bands. $\Lambda_2$ shows also the high energy peak at 
$\sim 1.5$ eV, which however does not directly couple to the RIXS process
since it contains two SO excitons residing on different sites. 
(ii) $\Lambda_3$ and $\Lambda_4$ ($J_{eff}=1/2$ e-h continuum) are located in 
the wide range of $0.5-1.5$ eV. This fact reveals that not only the lower but 
also the higher peak of $\sigma(\omega)$ is attributed to the e-h continuum of 
$J_{eff}=1/2$ states, contrary to previous 
interpretation based on a simple band picture~\cite{BJKim1}.
(iii) $\Lambda_3$ (NN e-h contribution) is depleted in the vicinity of the 
SO exciton $\Lambda_2$. The mixing of these modes is natural in view of the
nonzero overlap between quasi-degenerate $|ASDD\rangle \in \mathbf{H}_3$ and 
$|DQDD\rangle \in \mathbf{H}_2$ states [see Fig.~\ref{fig3}(a,b)]  
by virtue of intersite hoppings. 

The PES thus shows that the optically active e-h continuum of $J_{eff}=1/2$ 
band and the optically forbidden SO exciton~\cite{JKim} 
are located in the same energy range of $0.5-1.5$ eV, 
and there is considerable mutual interaction 
between these excitations due to the interband hopping between $J_{eff}=1/2$ 
and $J_{eff}=3/2$ states. Two peak structure in $\sigma(\omega)$ results from 
this unusual mixing among different excitations, which is reminiscent of the 
characteristic behavior of the Fano resonance. 

To conclude, we presented a unified description of magnetic couplings, optical
conductivity, and RIXS spectra in Sr$_2$IrO$_4$ within the model including 
strong SO coupling and ionic multiplet effects. The results obtained are 
consistent with the available experimental data. Since the SO splitting  
and the Mott-Hubbard charge gap are of similar scale, an unusual 
coupling between two different types of excitations -- the optically inactive 
SO exciton and the optically active e-h continuum -- is induced due to the 
interband mixing of $J_{eff}=1/2$ and $J_{eff}=3/2$ states. Although this effect 
is not very essential for NN magnetic interactions, it plays a crucial role 
in determining the shape of optical spectra: the e-h continuum is suppressed 
in the vicinity of the SO exciton inherent to iridates. The unusual Fano-type 
coupling between the rich ionic multiplet excitations and electronic continuum 
seems to be a characteristic feature of $J_{eff}=1/2$ Mott insulators. This
phenomenon is rooted to the atomic/band duality of 5d electrons and should 
therefore be generic to iridates with unusual hierarchy of energy scales.  

We thank G. Jackeli for useful discussions.
This work was supported by the NRF (No.2009-0079947).

\newpage

\setcounter{figure}{0}
\renewcommand{\thefigure}{S\arabic{figure}}

\section{{\it Supplemental Material}}

The gap value and peak positions in optical spectra are most sensitive to the
interaction parameters $U$ and $J_H$. At the same time, there is large
uncertainly concerning their values in iridates: current estimates for 
$U$ ($\sim 2.2$~eV~\cite{Mar11,Ari12}, 3 eV~\cite{Com12}) and for 
$J_H$ ($\sim 0.2$~eV~\cite{Ari12}, $0.3$~eV~\cite{Mar11}, 
$0.6$~eV~\cite{Com12,Mar88}) vary broadly. We therefore 
determine $U$ and $J_H$ values such that they give a reasonable agreement 
with the observed optical gap and peak positions, and, at the same
time, we cross-check them by calculating the magnetic exchange constants.

\begin{figure}[!b]
\centering
\includegraphics[width=8.0 cm]{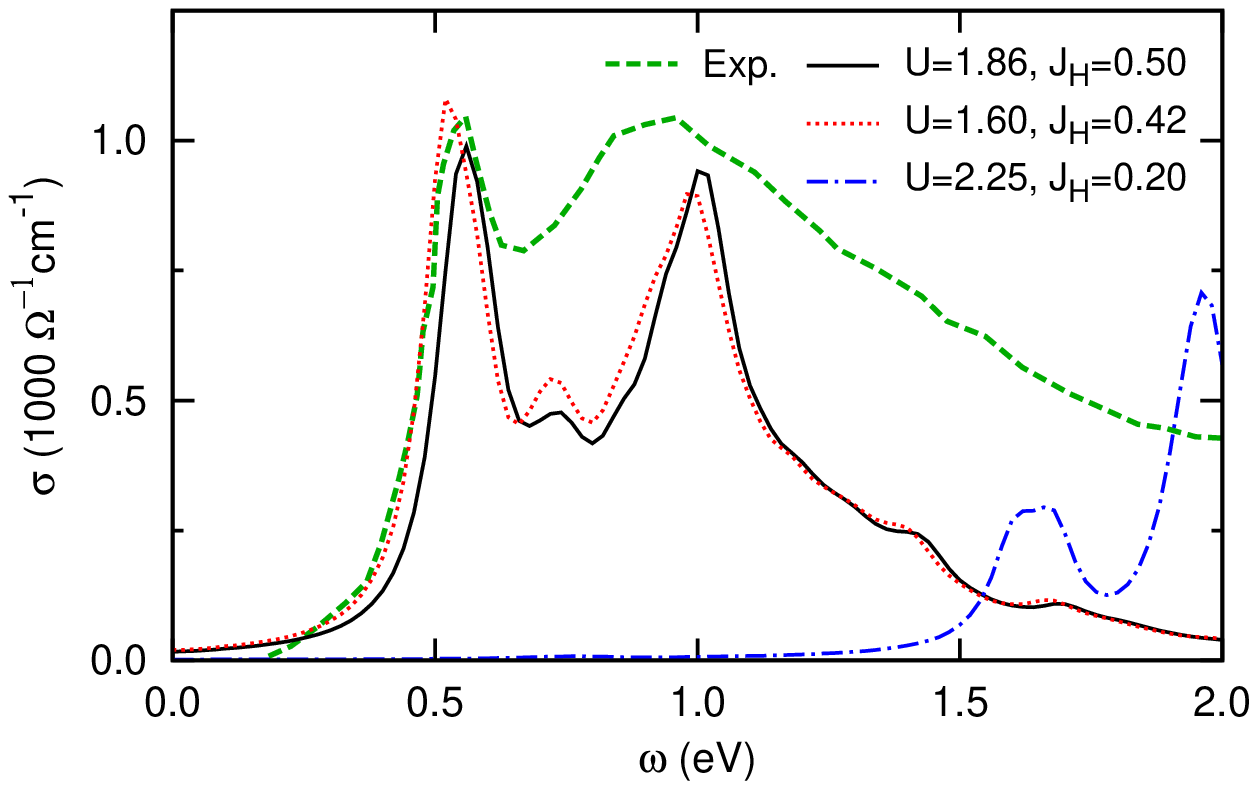}
\caption
{(color online)
Optical conductivity calculated with three different sets of $U$ and $J_H$ 
interaction parameters: $U=1.86$~eV and $J_H=0.50$~eV (solid line), 
$U=1.60$~eV and $J_H=0.42$~eV (dotted line), $U=2.25$~eV and $J_H=0.2$~eV 
(dashed-dotted line). Green dashed line represents the experimental data.
The Heisenberg exchange couplings $J$ calculated with these parameter sets 
are $J=76.8$~meV, $83.0$, and $37.8$ meV, respectively; the experimental 
value is $J \simeq 60$~meV~\cite{JKim}.
}
\label{Sfig}
\end{figure}

We found that $U\sim 2.2$~eV and $J_H\simeq 0.2-0.3$~eV suggested by
Refs.~\cite{Mar11,Ari12} result in too large optical gap (see a representative
curve in Fig.~\ref{Sfig}) and too small magnetic exchange constant $J$. 
Far better agreement with experiment is obtained when we slightly descrease 
$U$ and substantially increase $J_H$ (above 0.4~eV). Solid line in 
Fig.~\ref{Sfig} shows the result for $U=1.86$~eV and $J_H=0.5$~eV. 
As discussed in the main text, this parameter set provides also the magnetic 
interactions and spin canting angles that are in fair agreement with 
experiment. 

$U$ and $J_H$ can still be varied within the ranges 1.6-2.0 eV and 0.4-0.5 eV,
respectively, with acceptable fit results concerning the optical data. 
For instance, we show in Fig.~\ref{Sfig} the result for $U=1.60$~eV and 
$J_H=0.42$~eV (estimated as $J_H=3B+C$ using the Racah parameters $B$ and 
$C$ for an octahedrally coordinated Ir$^{4+}$ impurity~\cite{And76}). While 
these parameters provide optical spectra similar to that we found for 
$U=1.86$~eV and $J_H=0.5$~eV, the latter choice gives a magnetic coupling 
$J$ which is closer to the observed value.


\begin{thebibliography}{99}

\bibitem{Ima98} M. Imada, A. Fujimori, and Y. Tokura,  
   Rev. Mod. Phys. \textbf{70}, 1039 (1998). 

\bibitem{BJKim1} B.J. Kim, H. Jin, S.J. Moon, J.-Y. Kim, B.-G. Park,
C.S. Leem, J. Yu, T.W. Noh, C. Kim, S.-J. Oh, J.-H. Park, V. Durairaj,
G. Cao, and E. Rotenberg, 
   Phys. Rev. Lett. \textbf{101}, 076402 (2008).

\bibitem{SJMoon} S.J. Moon, H. Jin, K.W. Kim, W.S. Choi, Y.S. Lee, J. Yu,
G. Cao, A. Sumi, H. Funakubo, C. Bernhard, and T.W. Noh,
   Phys. Rev. Lett. \textbf{101}, 226402 (2008).

\bibitem{BJKim2} B.J. Kim, H. Ohsumi, T. Komesu, S. Sakai, T. Morita,
H. Takagi, and T. Arima,
   Science \textbf{323}, 1329 (2009).

\bibitem{Tak00} H. Takagi (unpublished). 

\bibitem{Singh} Y. Singh and P. Gegenwart, 
   Phys. Rev. B \textbf{82}, 064412 (2010).

\bibitem{Watanabe} H. Watanabe, T. Shirakawa, and S. Yunoki, 
   Phys. Rev. Lett. \textbf{105}, 216410 (2010).

\bibitem{Martins} C. Martins, M. Aichhorn, L. Vaugier, and S. Biermann,
   Phys. Rev. Lett. \textbf{107}, 266404 (2011).

\bibitem{Arita} R. Arita, J. Kune\v{s}, A.V. Kozhevnikov, A.G. Eguiluz, 
and M. Imada, 
   Phys. Rev. Lett. \textbf{108}, 086403 (2012).

\bibitem{JKim} J. Kim, D. Casa, M.H. Upton, T. Gog, Y.-J. Kim,
J.F. Mitchell, M. van Veenendaal, M. Daghofer, J. van den Brink,
G. Khaliullin, and B.J. Kim,  
   Phys. Rev. Lett. \textbf{108}, 177003 (2012).

\bibitem{Fuj12} S. Fujiyama, H. Ohsumi, T. Komesu, J. Matsuno, B.J. Kim, 
M. Takata, T. Arima, and H. Takagi, 
   Phys. Rev. Lett. \textbf{108}, 247212 (2012).

\bibitem{Jackeli} G. Jackeli and G. Khaliullin,
   Phys. Rev. Lett. \textbf{102}, 017205 (2009).

\bibitem{Ishii} K. Ishii, I. Jarrige, M. Yoshida, K. Ikeuchi,
J. Mizuki, K. Ohashi, T. Takayama, J. Matsuno, and H. Takagi,
   Phys. Rev. B \textbf{83}, 115121 (2011).

\bibitem{Kha04} G. Khaliullin, W. Koshibae, and S. Maekawa, 
   Phys. Rev. Lett. \textbf{93}, 176401 (2004).

\bibitem{Ament} L.J.P. Ament, G. Khaliullin, and J. van den Brink,
   Phys. Rev. B \textbf{84}, 020403(R) (2011).

\bibitem{SJMoon2} S.J. Moon, Hosub Jin, W.S. Choi, J.S. Lee,
S.S.A. Seo, J. Yu, G. Cao, T.W. Noh, and Y.S. Lee,
   Phys. Rev. B \textbf{80}, 195110 (2009).

\bibitem{Kov04} N.N. Kovaleva, A.V. Boris, C. Bernhard, A. Kulakov, 
A. Pimenov, A.M. Balbashov, G. Khaliullin, and B. Keimer, 
   Phys. Rev. Lett. \textbf{93}, 147204 (2004).

\bibitem{Khaliullin} G. Khaliullin, P. Horsch, and A.M. Ole\'{s},
   Phys. Rev. B \textbf{70}, 195103 (2004).

\bibitem{Kha05} G. Khaliullin,  
   Prog. Theor. Phys. Suppl. \textbf{160}, 155 (2005). 

\bibitem{Kan63} J. Kanamori, 
   Prog. Theor. Phys. \textbf{30}, 275 (1963). 

\bibitem{para1} In the limit of $10Dq\rightarrow\infty$, the energy splitting 
between $j_{eff}=1/2$ and $j_{eff}=3/2$ is $1.5\lambda$~\cite{Jackeli,Ament}. 
At finite $10Dq$ values, however, this splitting increases: for $10Dq=3.0$ eV, 
it is about $1.7\lambda$.

\bibitem{SJMoon3} S.J. Moon, M.W. Kim, K.W. Kim, Y.S. Lee, J.-Y. Kim,
J.-H. Park, B.J. Kim, S.-J. Oh, S. Nakatsuji, Y. Maeno,
I. Nagai, S.I. Ikeda, G. Cao, and T.W. Noh,
   Phys. Rev. B \textbf{74}, 113104 (2006).

\bibitem{Slater} J.C. Slater and G.F. Koster, 
   Phys. Rev. \textbf{94}, 1498 (1954).

\bibitem{BHKim} Beom Hyun Kim and B.I. Min, 
   New J. Phys. \textbf{13}, 073034 (2011).

\bibitem{Kat12} V.M. Katukuri, H. Stoll, J. van den Brink, and L. Hozoi, 
Phys. Rev. B \textbf{85}, 220402(R) (2012). 

\bibitem{Sch84} O.F. Schrimer, A.F\"orster, H. Hesse, M. W\"ohlecke, 
and S. Kapphan, 
J. Phys. C {\bf 17}, 1321 (1984).

\bibitem{SM} See {\it Supplemental Material} for details.

\bibitem{HJin} H. Jin, H. Jeong, T. Ozaki, and J. Yu,
   Phys. Rev. B \textbf{80}, 075112 (2009).

\bibitem{footnote} This is because an average value of dipole matrix element 
$|\langle ASDD|\hat J_c|DDDD\rangle|^2$ is about ten times larger than, e.g., 
$|\langle ATDD|\hat J_c|DDDD\rangle|^2$.

\end{thebibliography}

\begin{thebibliography}{99}

\bibitem[S1]{Mar11} C. Martins, M. Aichhorn, L. Vaugier, and S. Biermann, 
Phys. Rev. Lett. \textbf{107}, 266404 (2011).

\bibitem[S2]{Ari12} R. Arita, J. Kune\v{s}, A.V. Kozhevnikov, A.G. Eguiluz, 
and M. Imada, Phys. Rev. Lett. \textbf{108}, 086403 (2012).

\bibitem[S3]{Com12} R. Comin, G. Levy, B. Ludbrook, Z.-H. Zhu, C.N. Veenstra, 
J.A. Rosen, Yogesh Singh, P. Gegenwart, D. Stricker, J.N. Hancock, 
D. van der Marel, I.S. Elfimov, A. Damascelli, arXiv:1204.4471. 

\bibitem[S4]{Mar88} D. van der Marel and G.A. Sawatzky, 
Phys. Rev. B \textbf{37}, 10674 (1988).

\bibitem[S5]{JKim} J. Kim, D. Casa, M.H. Upton, T. Gog, Y.-J. Kim, J.F. Mitchell, 
M. van Veenendaal, M. Daghofer, J. van den Brink, G. Khaliullin, and B.J. Kim, 
Phys. Rev. Lett. \textbf{108}, 177003 (2012).

\bibitem[S6]{And76}  B. Andlauer, J. Schneider, and W. Tolksdorf, 
Phys. Stat. Sol. B \textbf{73}, 533 (1976).  

\end{thebibliography}
\end{document}